\documentclass[]{spie}  

\usepackage{amsmath,amsfonts,amssymb}
\usepackage{graphicx}
\usepackage[colorlinks=true, allcolors=blue]{hyperref}

\title{Characterization of LBT atmospheric and turbulence conditions in the context of ALTA project}

\author[a]{Turchi, A.}
\author[a]{Masciadri, E.}
\author[b]{Veillet, C.}
\affil[a]{INAF-Osservatorio Astrofisico di Arcetri, L.go Enrico Fermi 5, Firenze, Italy}
\affil[b]{Large Binocular Telescope Observatory, 933 North Cherry Avenue, Tucson, AZ 85721, USA}

\authorinfo{Send correspondence to Alessio Turchi - e-mail: alessio.turchi@inaf.it}

\begin{document} 
\maketitle

\begin{abstract}
ALTA project has been active since 2016, providing, at LBT observatory site, forecasts of atmospheric parameters, such as temperature, wind speed and direction, relative humidity and precipitable water vapor, and optical turbulence parameters, such as seeing, wavefront coherence time and isoplanatic angle with the final goal to support nightly the science operation of the LBT. Besides to the forecasts, during the years ALTA has been collecting statistics on the atmospheric conditions which can be used to draw a very accurate characterization of the climatology of the telescope site located on top of Mount Graham, Arizona. Such characterization can be used both for the optimization and calibration of the forecast model and as a reference for a model validation. The climatology of these parameters is conceived to be a further output of ALTA that will be upgraded on the website with time and it will be able to put in evidence trends at short as well as long time scales. In this contribution we present a climatological description of all the atmospheric parameters relevant for ground-based astronomy in order to provide to the scientific community a robust reference of the observing conditions at LBT. The study is performed using on-site measurements provided by DIMM and atmospheric sensors over several years and made available in the telescope telemetry data.
\end{abstract}

\keywords{site characterization,atmosphere,optical turbulence,climatology}

\section{INTRODUCTION}
\label{sec:intro}  
The Large Binocular Telescope (LBT), which is part of the Mount Graham International Observatory (MGIO), is located in southeastern Arizona, at Mount Graham. The telescope is founded by an international collaboration of institutions grouped in the LBTO foundation, that includes different USA universities and research institutes (University of Arizona, University of Minnesota, University of Notre Dame, University of Virginia, Ohio State University and the Research Corporation for Science Advancement), the Istituto Nazionale di Astrofisica (INAF, the Italian National Institute for Astrophysics) and the LBT Beteiligungsgesellschaft in Germany (Max Planck Institute for Astronomy, Landessternwarte, Leibniz Institute for Astrophysics Potsdam, Max Planck Institute for Extraterrestrial Physics and Max Planck Institute for Radio Astronomy).\\
The LBT is a unique installation since it has two twin 8.4 m telescopes mounted on the same structure in a binocular setup, which can operate together in interferometric configuration (with a combined baseine of 22.8 m), or separately, even in dual configuration optical/interferometric [\citenum{hill2012}].\\
The LBT houses advanced Adaptive Optics (AO) systems and, since the last decades, has pioneered the field of extreme-AO with the FLAO system [\citenum{esposito2012}], one of the first AO systems to operate with a Pyramid Wavefront Sensor (WFS), recently updated to SOUL [\citenum{pinna2016}], that feeds the camera LUCI [\citenum{seifert2003}], the Large Binocular Telescope Interferometer (LBTI) [\citenum{hinz2012}], and the upcoming extreme-AO SHARK-NIR [\citenum{farinato2016}] and SHARK-VIS [\citenum{pedichini2016}]. The telescope has also an advanced Laser Guide Star facility to support Gound-Layer AO (GLAO) [\citenum{rabien2019}].\\

The LBT is currently one of the few telescopes installation having a dedicated automatic forecast system for the prediction of the Optical Turbulence (OT) astroclimatic parameters (seeing ($\epsilon$), isoplanatic angle ($\theta_0$), wavefront coherence time ($\tau_0$)), and other atmospheric parameters relevant for ground based astronomy (wind speed (WS) and direction (WD), temperature (T), relative humidity (RH), precipitable water vapor (PWV)). Such a system  is the  ALTA Center \footnote{\href{http://alta.arcetri.inaf.it}{http://alta.arcetri.inaf.it}}. ALTA has been validated for predicting reliable atmospheric parameters[\citenum{turchi2017}]. This system operates serving both long-term forecasts (typically available at the beginning of the afternoon for the coming night i.e. 6h-15h in advance) and short-term forecasts (1h and 2h in advance) [\citenum{masciadri2020}]. Short term forecasts are updated each hour during the night and allows for very high accuracy, by making use of atmospheric measurements provided by the sensors installed on the telescope.\\

The forecast of the atmosphere and the OT is fundamental for both observation planning, by allowing correct matching between the observing program and the needed atmospheric conditions, and the optimization of the performance of the AO systems operating at LBT, whose performance depends on the atmospheric and OT conditions ($\epsilon$, $\theta_0$ and $\tau_0$) [\citenum{pedichini2016}, \citenum{rabien2019}].\\

ALTA supports the daily operations at the telescope by allowing a highly accurate evaluation of the atmospheric conditions shortly before or even during the operation, however a long-term characterization of the LBT atmospheric conditions (climatology) is equally of fundamental importance. On one side, long-term (months ahead) telescope observation planning can't rely alone on ALTA forecast, and a knowledge of the seasonal variations of the atmosphere can help to carefully evaluate the schedule of long-term projects. AO development is also impacted by the characterization of the climatology of the site, since the design have to adopt different decisions based on the typical (median) operating conditions of the observing site (i.e. vertical distribution of the turbulence, wind conditions, typical values of OT parameters...). Finally the ALTA forecast itself is impacted by the climatological characterization of the site. ALTA is based on the Astro-Meso-NH code [\citenum{masciadri1999}], which ideally requires a calibration of the OT on the specific forecast site by using both the OT vertical profiles (provided by a profiler such as STEREO-SCIDAR) and integrated seeing measurements (provided by a DIMM), with the largest possible available statistic [\citenum{masciadri2019}].\\

In this contribution we aim to provide a climatological description of the LBT site by using the measurements stored in the LBT telemetry and obtained with the instruments available at the telescope site. To this aim we considered the period from 2013 to 2021 (9 years), which is the largest available at present from the historical archive of the telemetry data.\\
In section \ref{datasource} we will describe the measurement instruments used for this analysis and their setup.\\
In section \ref{climatolory} we will show the climatological characterization of the LBT site, both in each single year considered in our reference period and as a global characterization over the whole 9 years.\\
Finally in section \ref{conclusions} we will draw the conclusions.\\

\section{Atmospheric and OT measurements}
\label{datasource}
\begin{figure}
\centering
\includegraphics[width=1.0\textwidth]{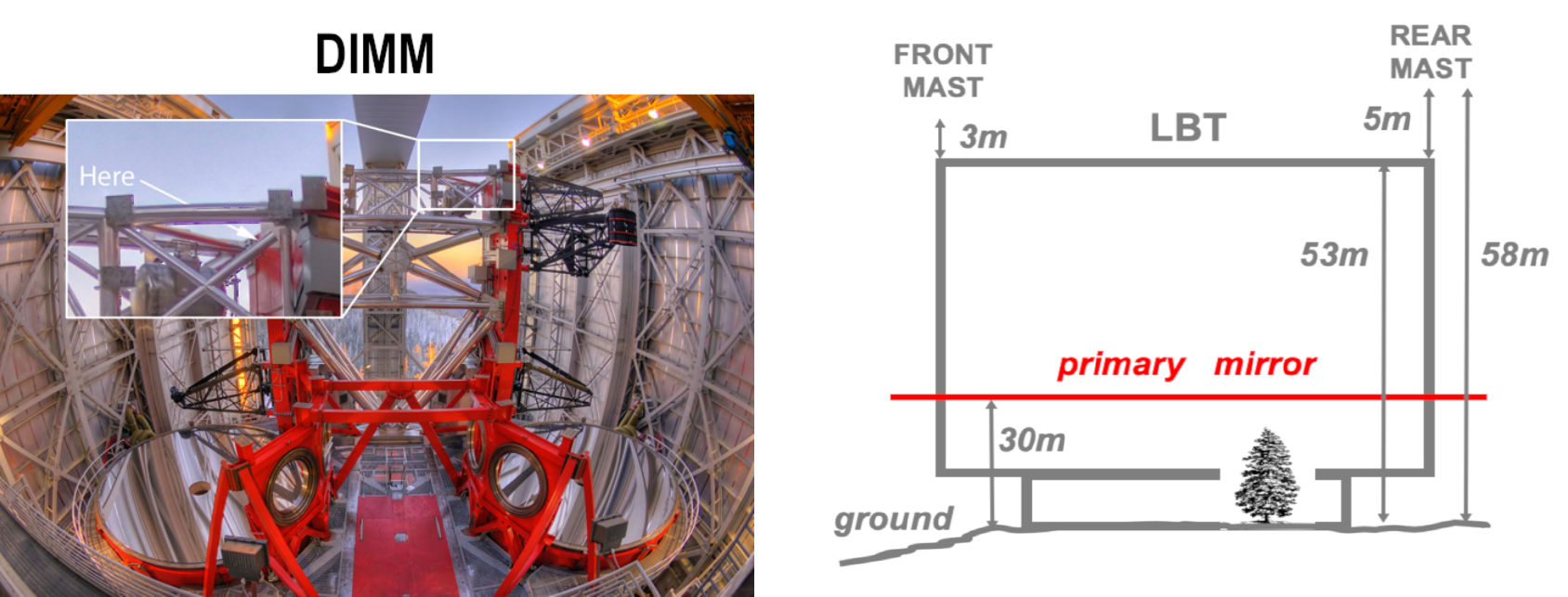}
\caption{Left: Location of the DIMM inside the LBT dome. Right: Location of the two weather masts on the LBT roof.}
\label{fig:mast}
\end{figure}

The LBT has different measurement sources for the atmospheric and OT data. Seeing measurements are produced by a DIMM, which is placed inside the dome on the telescope structure (see Fig. \ref{fig:mast}, left panel). This means that the DIMM value also takes into account some of the dome seeing, which is the turbulence produced inside the telescope dome itself. This is a double edged feature that has both drawbacks (the inability to correctly disentangle the contribution to the total seeing coming from the atmosphere and the temperature gradients inside the dome) and advantages. Such a location has the advantage to provide a turbulence estimate more realistic with respect to what happens on the CCD of the instrument at the focus of the telescope. This latter aspects means that Astro-Meso-NH and short-term forecasts, based on autoregression techniques (AR) that make use of real-time data, can both be calibrated (the second one in an automatic way due to the AR process) to take into account this additional source of OT.\\

\begin{figure}
\centering
\includegraphics[width=1.0\textwidth]{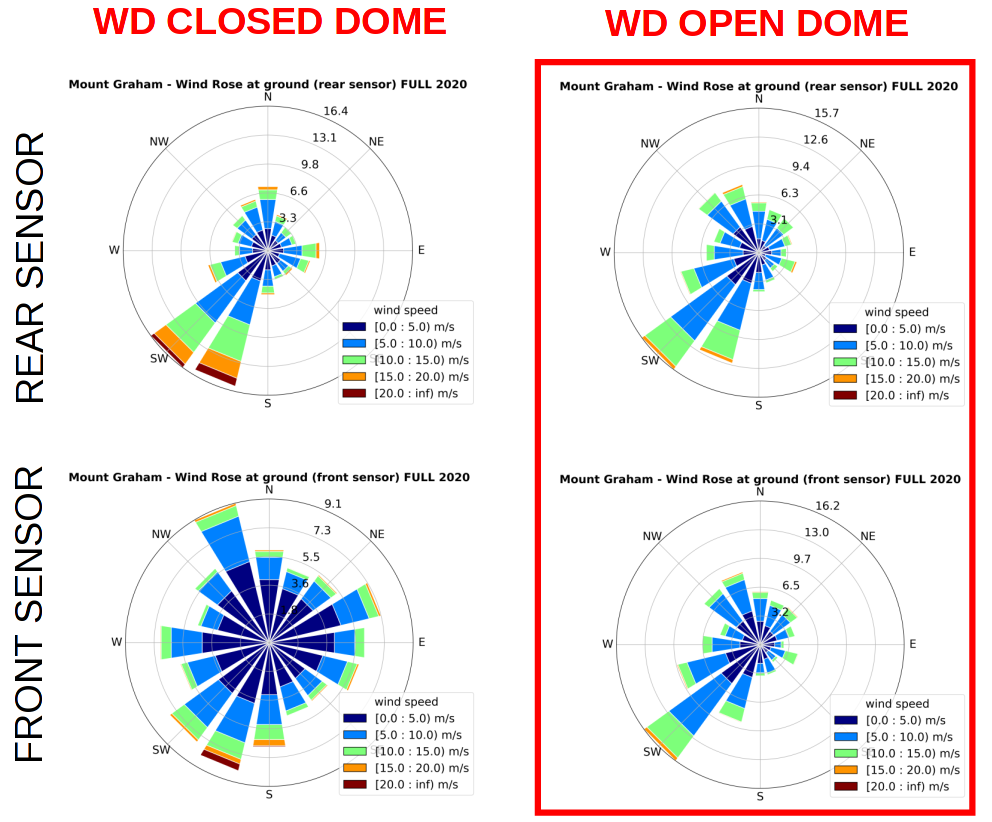}
\caption{Comparison of the wind rose computed on 2020 data between front and rear sensors. On the right column the wind rose with closed dome, on the left the wind rose computed only when the dome is open. We can see that the two sensors have an almost identical distribution in open configuration, while the closed dome configuration causes the front sensor to produce a clearly unrealistic wind distribution.}
\label{fig:wdopen}
\end{figure}

Atmospheric data used in this climatological characterization, that is temperature, relative humidity, wind speed and direction, are obtained by the weather masts placed 5 meters above the roof of the dome, in the ``rear'' position (see Fig. \ref{fig:mast}, right panel). There is also  another mast in the ``front'' position, placed 3 m above the dome, measuring only wind speed and direction. Measurements from both masts revealed very useful to retrieve reliable measurements on the wind. When the sensors are placed on the top of the dome it is always suitable to have at least a couple of sensors to discriminate from spurious effects due to the relative position of the building with respect to the main wind direction. We observed that wind data from the front and the rear sensor are similar only when the dome is open, while the front measurements have an unrealistic distribution when the dome is in closed position (see Fig. \ref{fig:wdopen}). Due to this fact, pending more in-situ investigations on the potential problems of the front sensor, we decided, for this contribution, to consider only wind measurements coming from the rear sensor.\\

We selected 9 years of data for this study, from 2013 to 2021, which is the largest period available in the telemetry archive of LBT and stored in hdf5 format. In the telemetry data, each measurement is associated to an ``alive'' flag that communicates if the corresponding instrument is operating correctly, so we selected all data based on this condition. Finally we discarded few residual non-physical values (e.g. measurements of seeing, wind speed, relative humidity lower than 0, or relative humidity values greater than 100).\\

\section{LBT climatology}
\label{climatolory}
We report here the cumulative distributions for the atmospheric and OT parameters considered in this analysis, either through the whole cumulative distributions over the full 9 years, or the individual yearly cumulative distributions. Each plot shows the global distribution and the winter (October-March) and summer (April-September) ones, to show the seasonal variation.\\
We also computed the mean value of the parameters, the first and last tertile of the distribution, each reported in the figures. We report the results of the above values computed over all the 9 years considered in this study in Table \ref{tab:tabdata}.\\

\begin{table}[ht]
\caption{Median, first and last tertile of the distribution of the atmospheric and OT parameters (except wind direction). The values are computed over the whole 9 years (2013-2021) considered in the present study. In the parentesis we report the summer/winter values.}
\label{tab:tabdata}
\begin{center}       
\begin{tabular}{|l|c|c|c|} 
\hline
{\bf Parameter} & {\bf Median} & {\bf First Tertile} & {\bf Last Tertile} \\
\hline
Seeing ('') & 1.0 (0.98/1.00) & 0.89 (0.88/0.89) & 1.13 (1.10/1.15) \\
\hline
Temperature (°C) & 5 (10.3/1.3) & 1.8 (8.0/0.9) & 8.7 (11.9/3.4) \\
\hline
Relative Humidity (\%) & 43.8 (47.9/38.6) & 30.7 (35.8/25.8) & 60.1 (62.7/56.6) \\
\hline
Wind Speed (m/s) & 6.4 (5.3/7.7) & 4.6 (3.8/5.6) & 8.6 (7.0/10.0) \\
\hline
\end{tabular}
\end{center}
\end{table}

\begin{figure}
\centering
\includegraphics[width=0.6\textwidth]{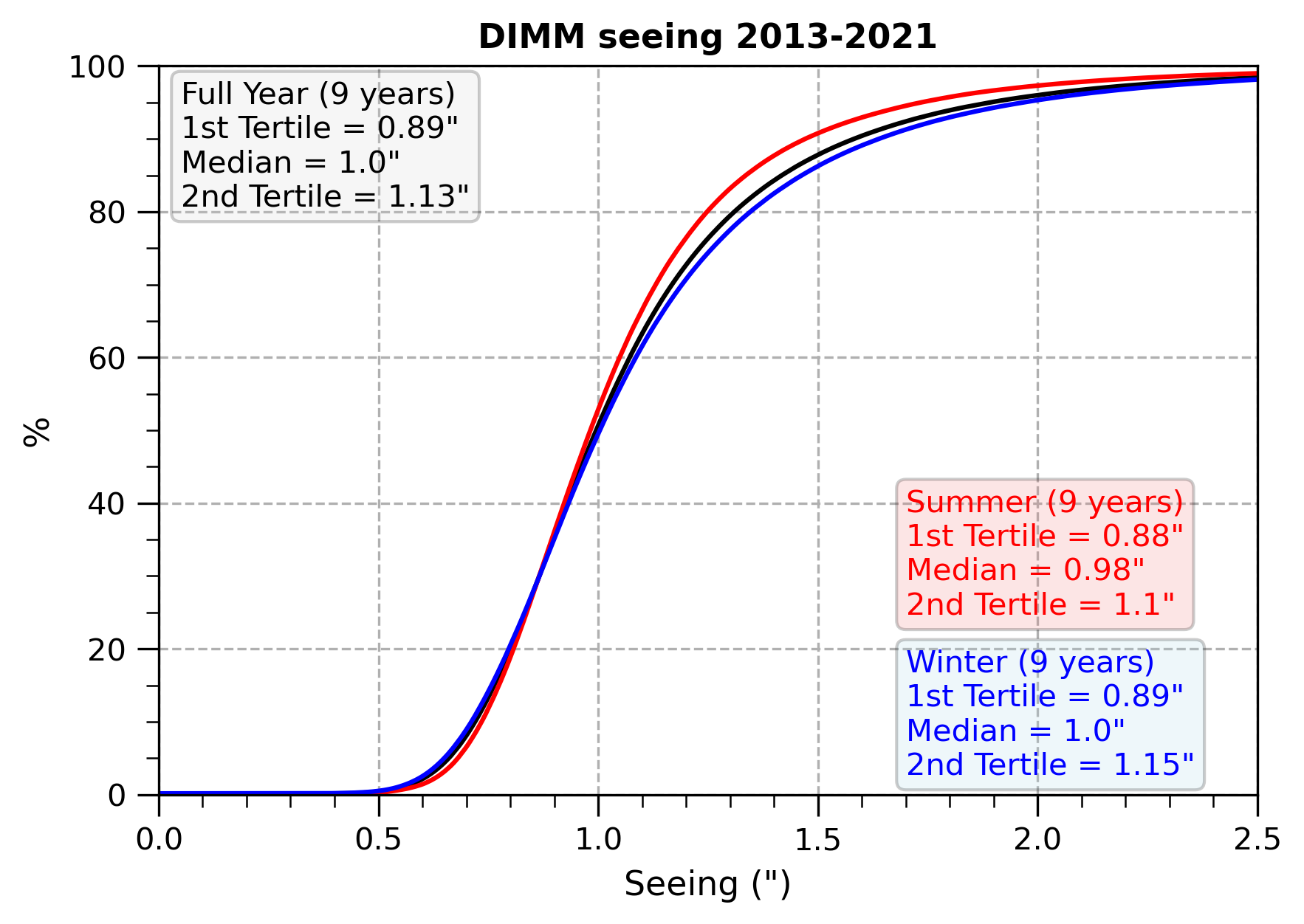}
\caption{Seeing cumulative distributions for all the years 2013-2021. Red and blue curves are relative to the summer and winter seasons respectively. The black curve is the full distribution.}
\label{fig:dimmcum}
\end{figure}

\begin{figure}
\centering
\includegraphics[width=1.0\textwidth]{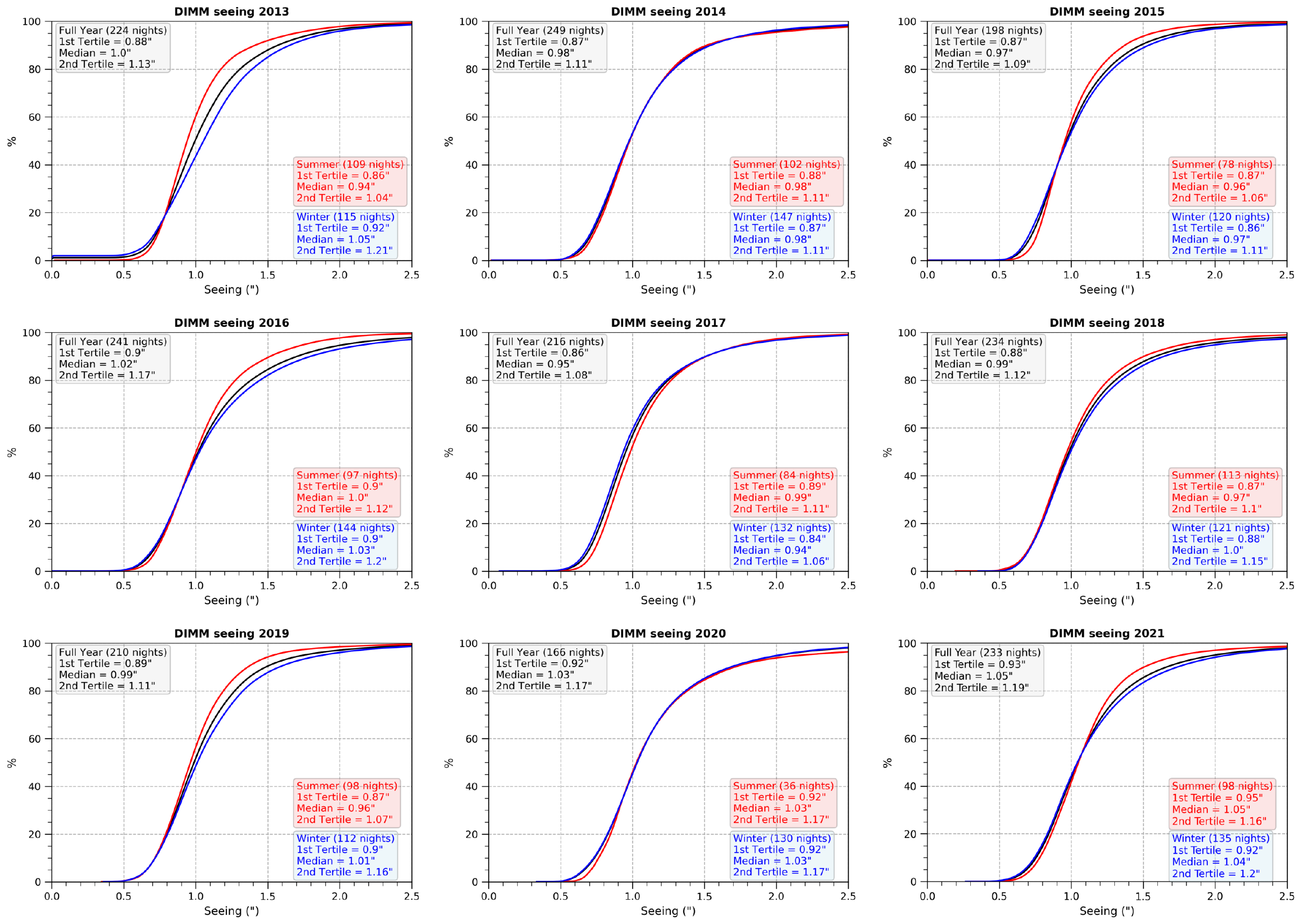}
\caption{Seeing cumulative distributions in each individual year 2013-2021. Red and blue curves are relative to the summer and winter seasons respectively. The black curve is the full distribution.}
\label{fig:seeall}
\end{figure}

Fig. \ref{fig:dimmcum} shows the cumulative distribution of seeing measurements in the 9 years period, while Fig. \ref{fig:seeall} reports the distributions in the individual years. From the data we evince that the LBT site is extremely stable, with a median seeing of $\sim$1.0” (see Table \ref{tab:tabdata}) and minimal variations of the value through the years. Also seasonal variation seems not to have a significant impact, since most years show an almost identical distribution in summer and winter. Winter has a slight more abundance of high seeing values, however a difference is present only in the 20\% higher quantile. Unique exception the year 2017, when there was a large fire during summer at Mount Graham that almost reached the LBT telescope, in which the seeing in summer was higher than in winter.\\

\begin{figure}
\centering
\includegraphics[width=0.6\textwidth]{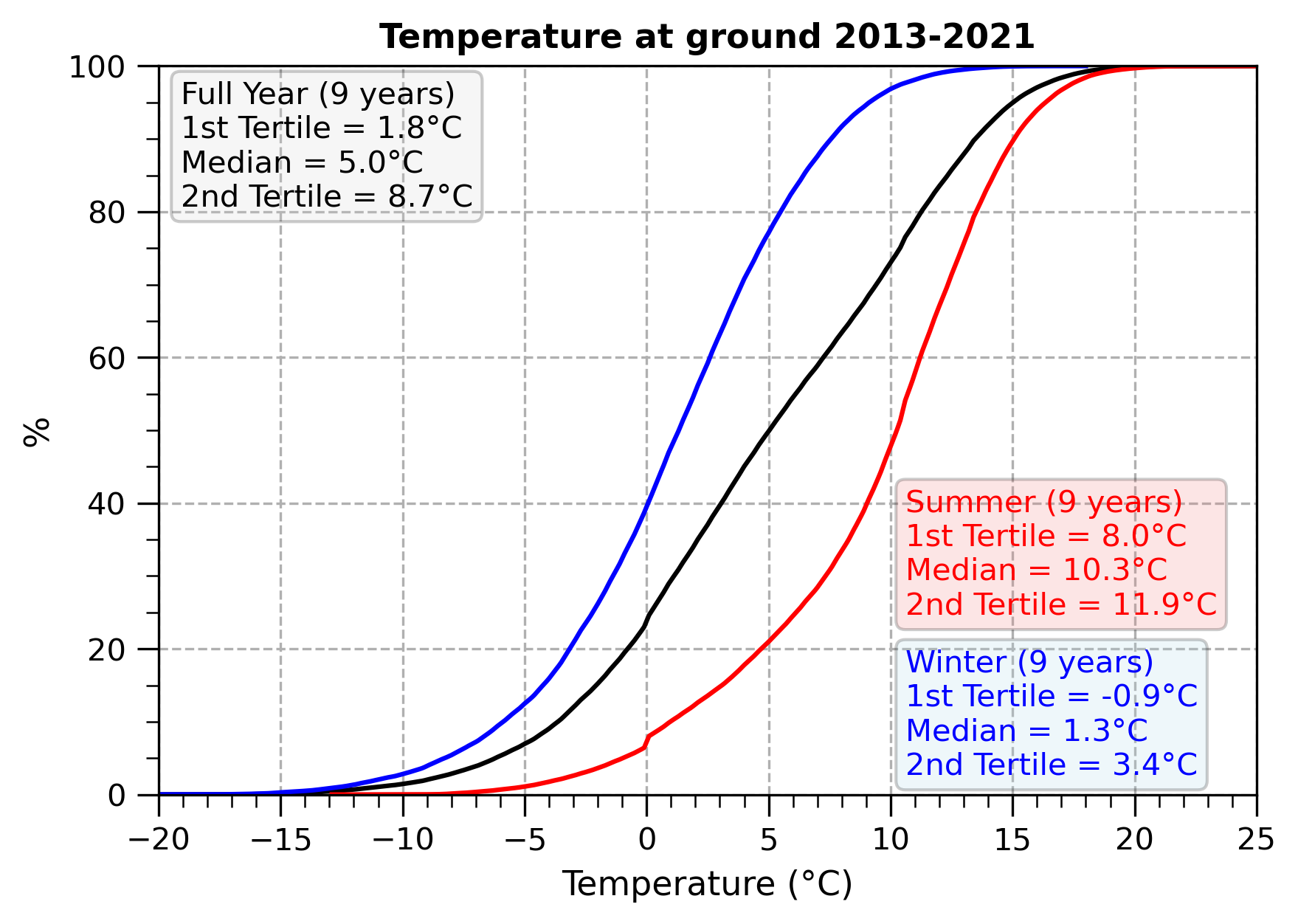}
\caption{Temperature cumulative distributions for all the years 2013-2021. Red and blue curves are relative to the summer and winter seasons respectively. The black curve is the full distribution.}
\label{fig:tempcum}
\end{figure}

\begin{figure}
\centering
\includegraphics[width=1.0\textwidth]{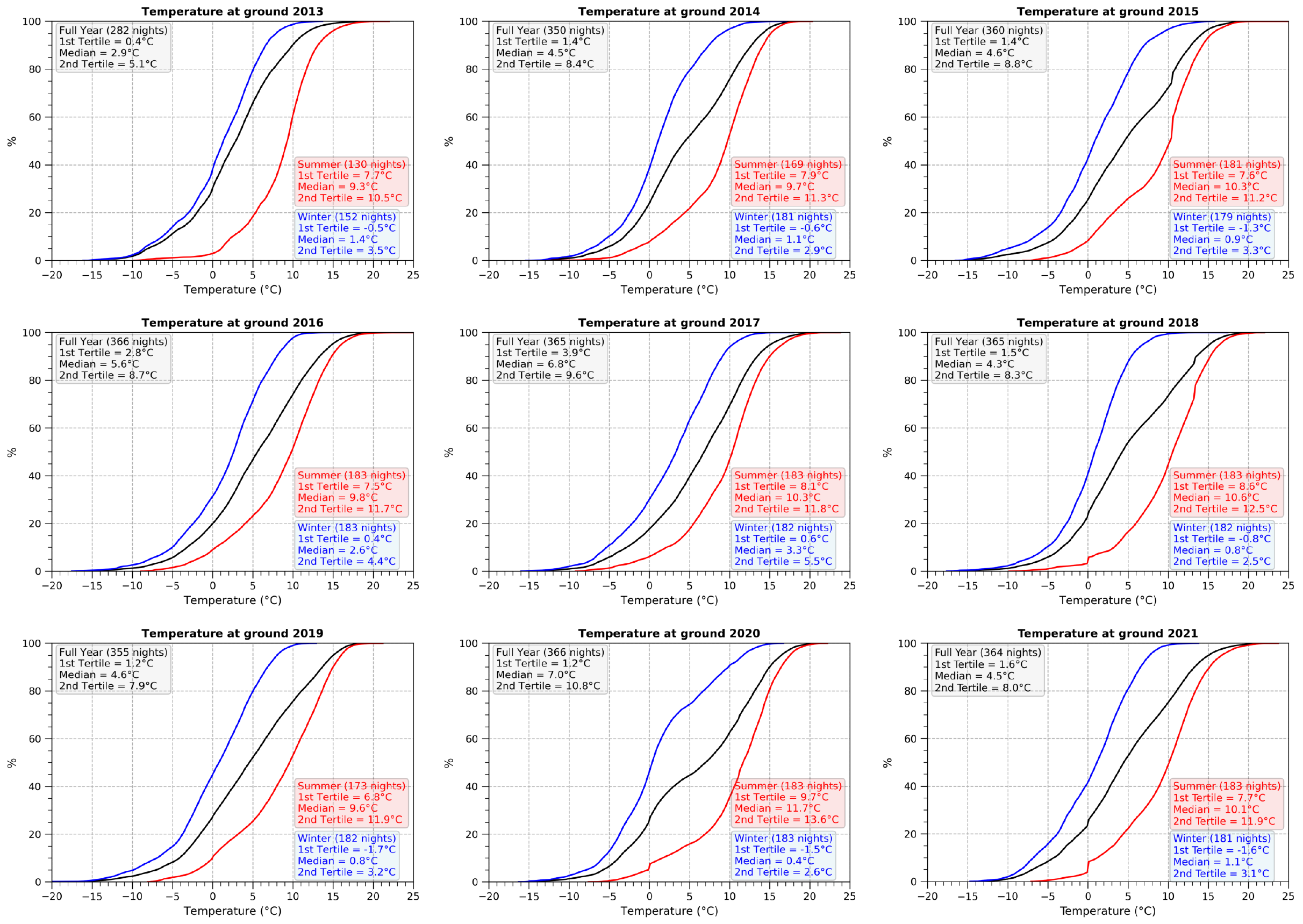}
\caption{Temperature cumulative distributions in each individual year 2013-2021. Red and blue curves are relative to the summer and winter seasons respectively. The black curve is the full distribution.}
\label{fig:tempall}
\end{figure}

Fig. \ref{fig:tempcum} shows the cumulative distribution of the temperature at LBT, with each year available in Fig. \ref{fig:tempall}, showing the expected seasonal variation. The median temperature on the site is around 5 °C, changing from 2 °C to 10 °C from winter to summer. The amount of time interested by temperatures below 0 °C is around 40\% in winter and less than 10\% in summer. Yearly variation show no evident pattern in the values of the median and the tertiles.\\

\begin{figure}
\centering
\includegraphics[width=0.6\textwidth]{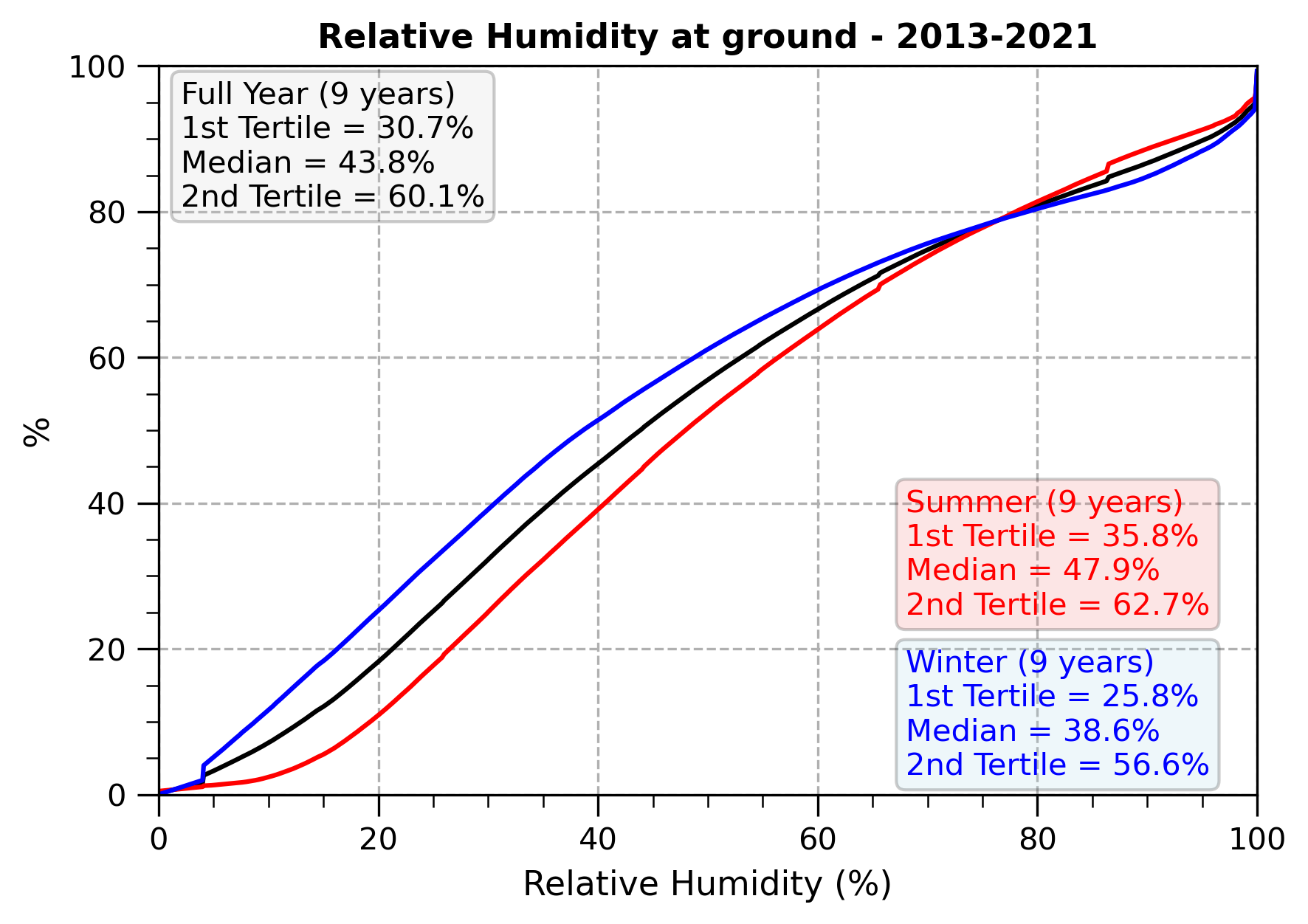}
\caption{Relative Humidity cumulative distributions for all the years 2013-2021. Red and blue curves are relative to the summer and winter seasons respectively. The black curve is the full distribution.}
\label{fig:rhcum}
\end{figure}

\begin{figure}
\centering
\includegraphics[width=1.0\textwidth]{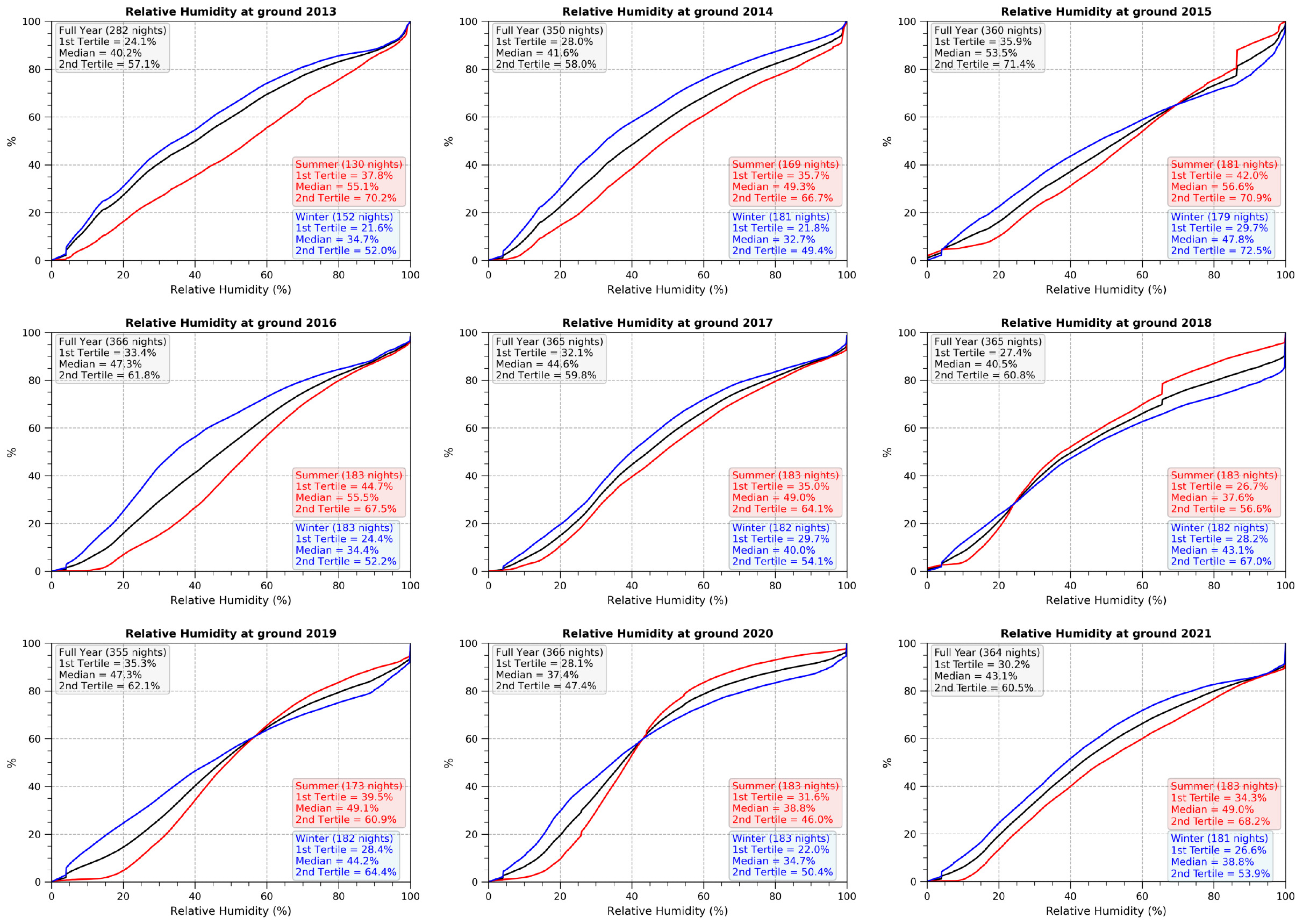}
\caption{Relative Humidity cumulative distributions in each individual year 2013-2021. Red and blue curves are relative to the summer and winter seasons respectively. The black curve is the full distribution.}
\label{fig:rhall}
\end{figure}

Fig. \ref{fig:rhcum} shows the cumulative distribution of relative humidity, with individual years in Fig. \ref{fig:rhall}. LBT has a median RH value of 44\%, with equally distributed 20\% dry nights (RH $<$ 20\%) and 20\% wet nights (with RH $>$ 80\%), where the telescope operations may be impacted by drops forming on the primary mirror. Also the percentage of dry atmosphere makes it suitable for infrared observations, which are mainly affected by the water contend in the atmosphere. While there is no monitor for the total water content (the precipitable water vapor), the same quantity is forecasted by the ALTA project. The typical water content profile is decreasing until around 10 km above ground (when water content becomes negligible), thus a dry atmosphere at ground means that the water profile should, on average, be low as well. From the yearly variation we see that recent years (specifically 2018, 2019 and 2020) are showing summer seasons that are more extreme, showing less dry events and less wet events, with a steeper cumulative distribution (contrary to the general tendency in the past years 2013-2017). It's too early to understand if this is a general tendency, however this could be a sign of a climate change that is worth monitoring in the future.\\

\begin{figure}
\centering
\includegraphics[width=0.6\textwidth]{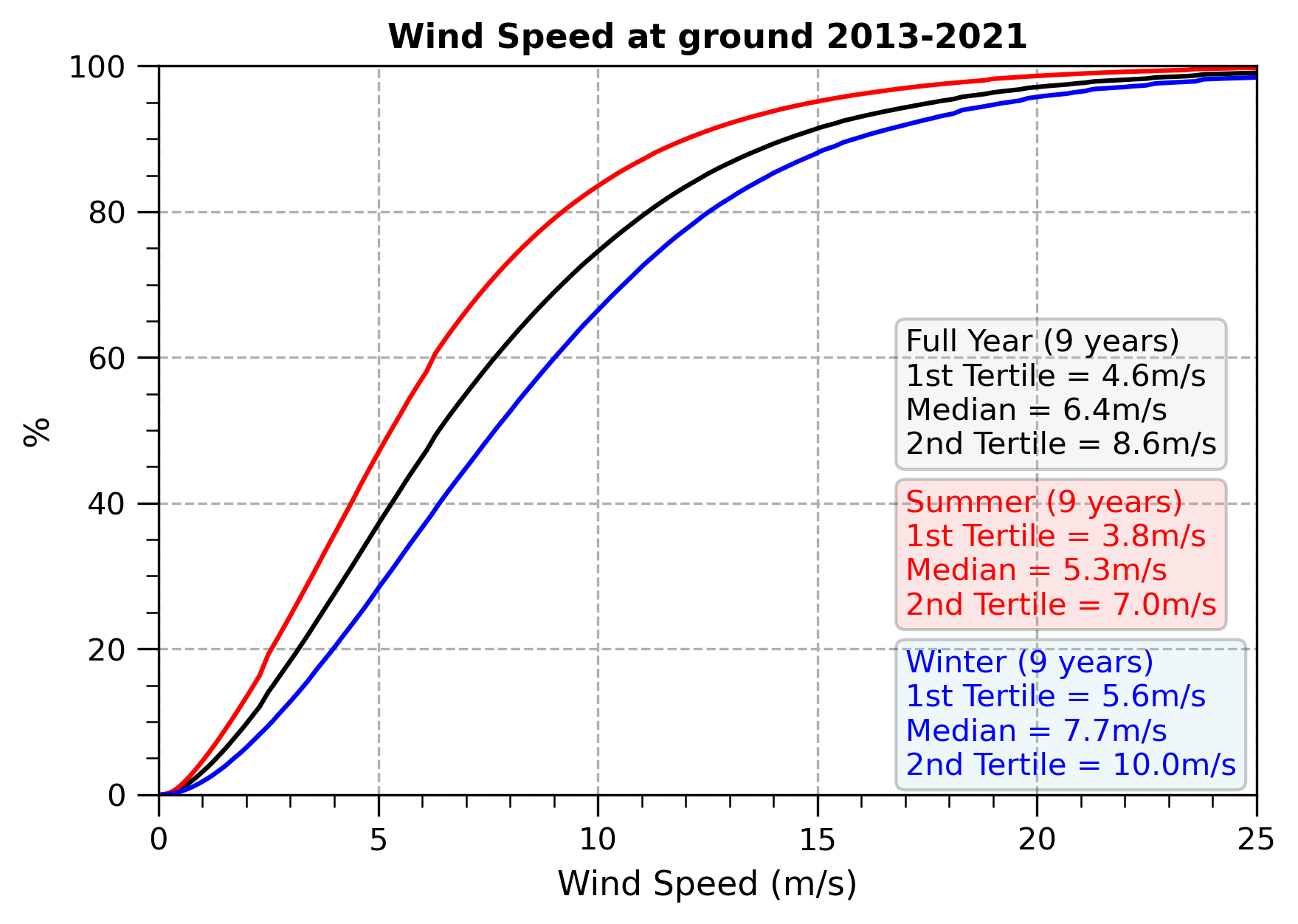}
\caption{Wind Speed cumulative distributions for all the years 2013-2021. Red and llue curves are relative to the summer and winter seasons respectively. The black curve is the full distribution.}
\label{fig:wscum}
\end{figure}

\begin{figure}
\centering
\includegraphics[width=1.0\textwidth]{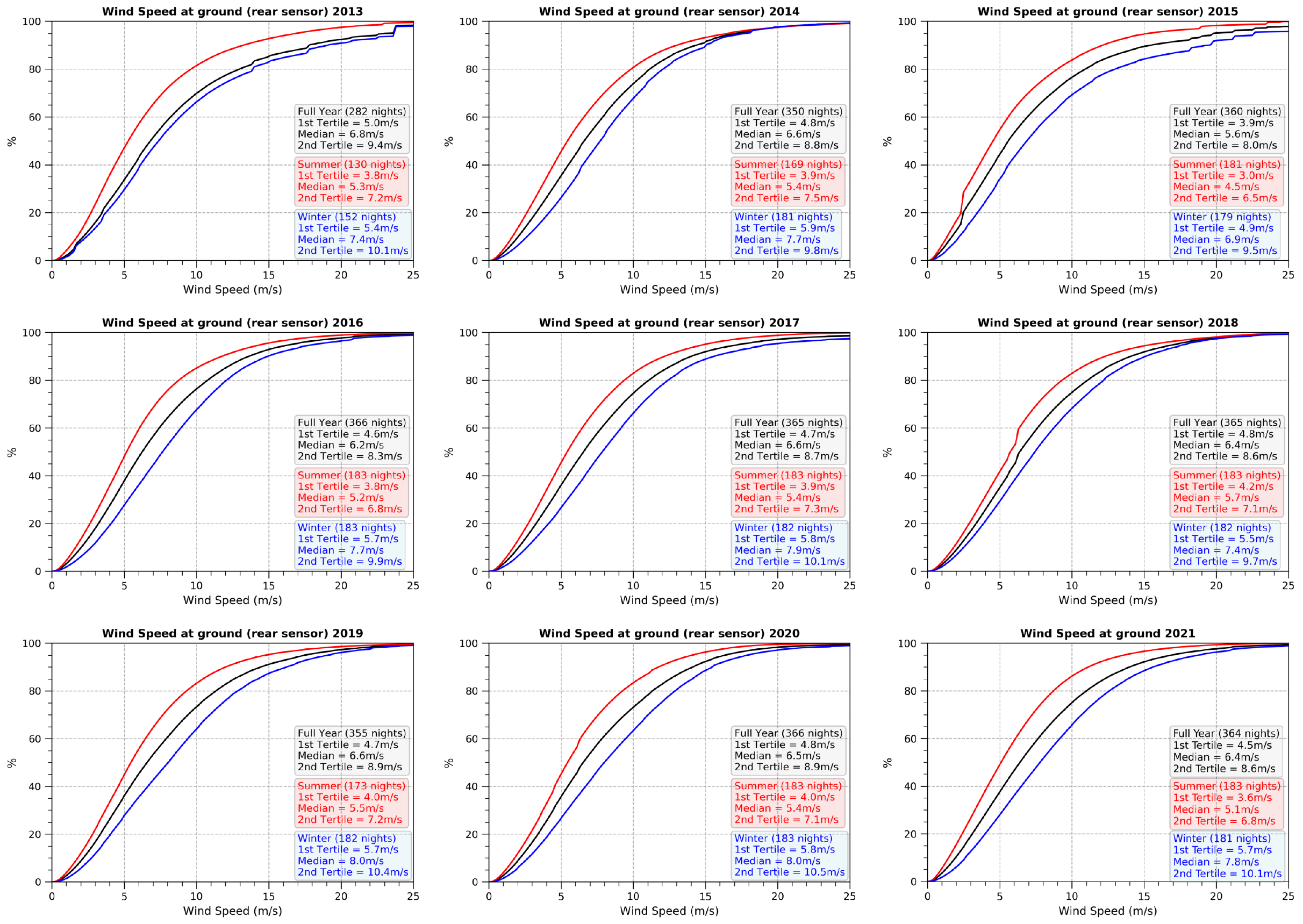}
\caption{Wind Speed cumulative distributions in each individual year 2013-2021. Red and blue curves are relative to the summer and winter seasons respectively. The black curve is the full distribution.}
\label{fig:wsall}
\end{figure}

Fig. \ref{fig:wscum} shows the distribution of the wind speed, with individual years reported in Fig. \ref{fig:wsall}. The median value of the wind speed is 6.4 ms$^{-1}$, showing a prevalence of strong winds during the winter season, with almost no yearly variation. Globally really strong winds ($\ge$ 15 m/s), that could force telescope operations to stop at all, are relatively rare (10-15\% of the time).\\

\begin{figure}
\centering
\includegraphics[width=0.7\textwidth]{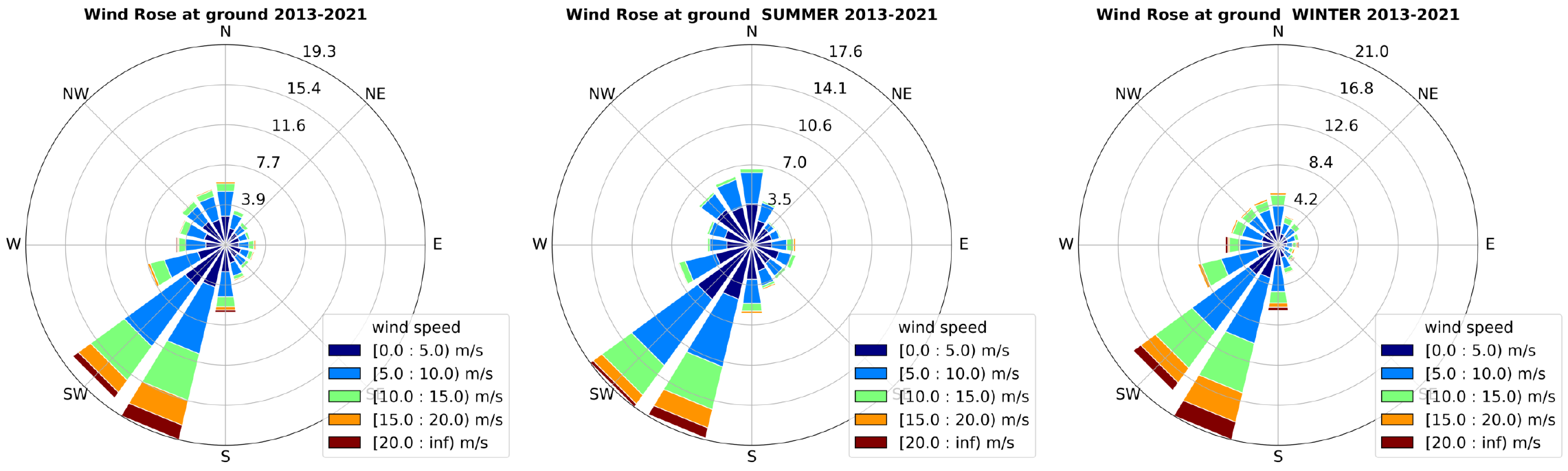}
\caption{Wind rose for all the years 2013-2021. Left figure refers to the whole wind distribution, while center and right figure are the wind roses computed over summer and winter respectively. Different colors correspond to different wind speed strengths.}
\label{fig:wdcum}
\end{figure}

\begin{figure}
\centering
\includegraphics[width=1.0\textwidth]{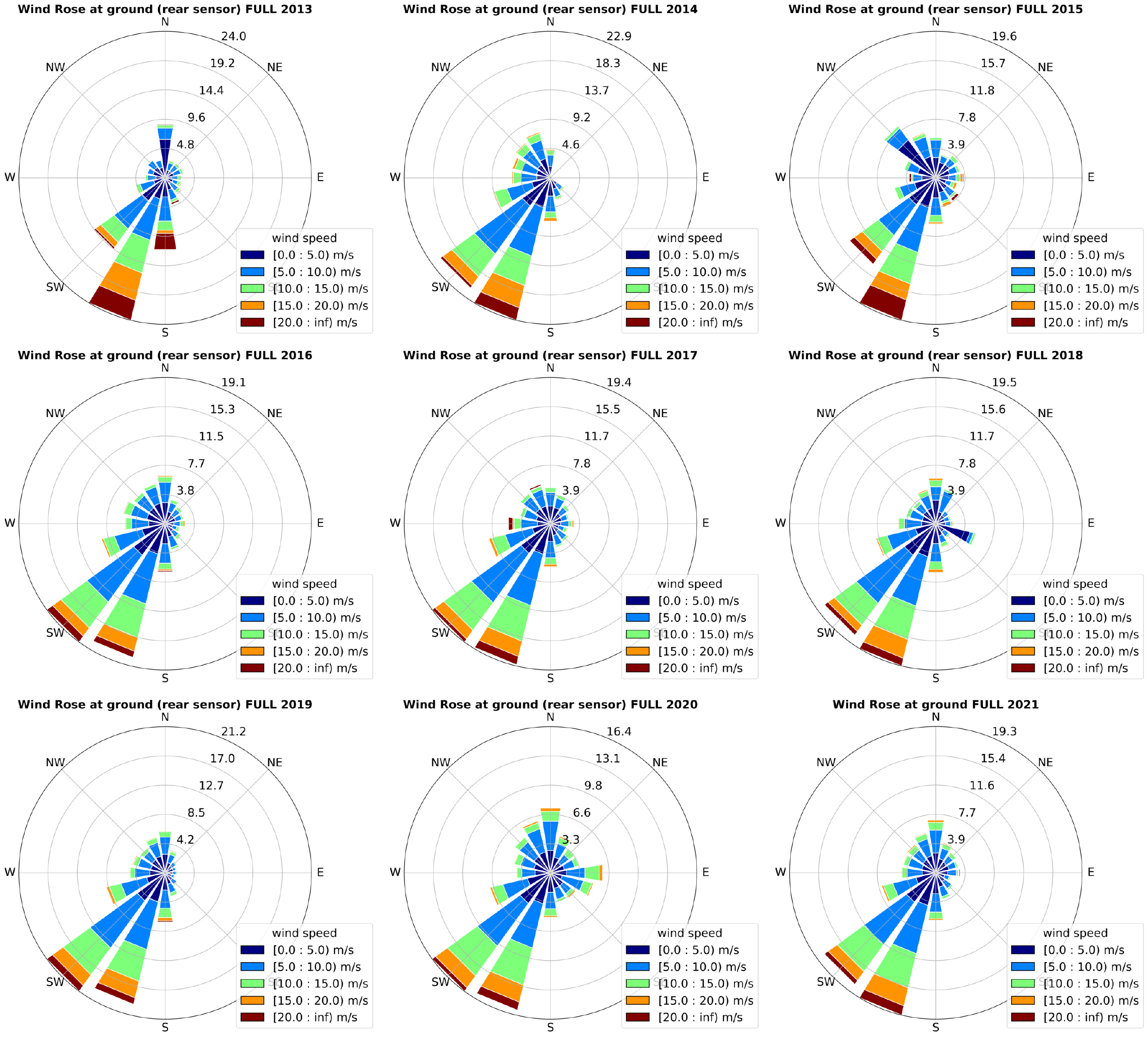}
\caption{Wind Rose for the individual years 2013-2021. This is the whole wind distribution for each year. Different colors correspond to different wind speed strengths.}
\label{fig:wdall}
\end{figure}

\begin{figure}
\centering
\includegraphics[width=1.0\textwidth]{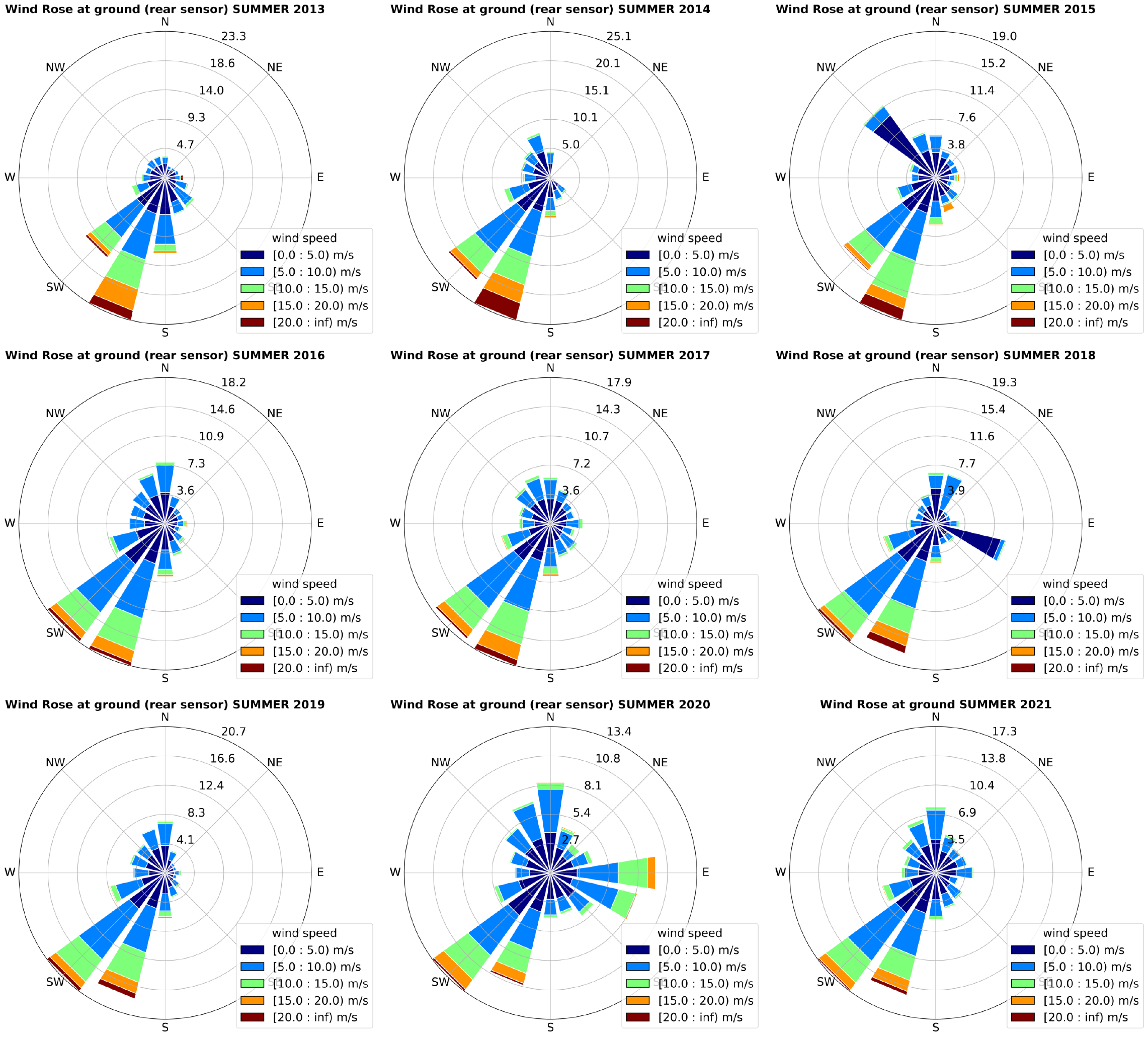}
\caption{Wind Rose for the individual years 2013-2021. This is relative to the summer season of each year. Different colors correspond to different wind speed strengths.}
\label{fig:wdsum}
\end{figure}

\begin{figure}
\centering
\includegraphics[width=1.0\textwidth]{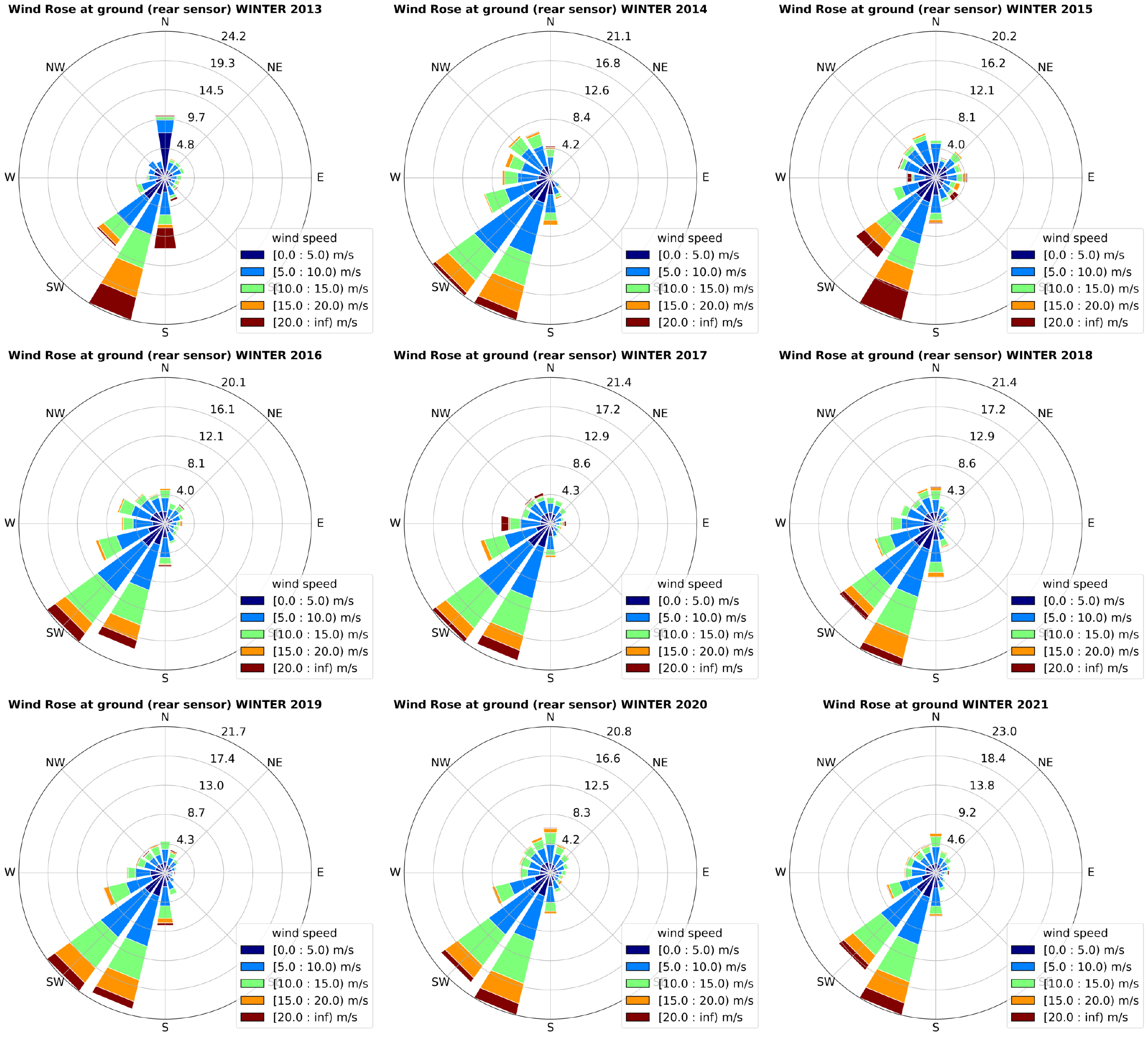}
\caption{Wind Rose for the individual years 2013-2021. This is relative to the winter season of each year. Different colors correspond to different wind speed strengths.}
\label{fig:wdwin}
\end{figure}

In Fig. \ref{fig:wdcum} we report the wind rose computed over the 9 years and divided into the full wind rose on the whole sample (left) and the two partial roses computed over the summer and winter seasons. We see that, independently on the season, most of time the wind flows from the SW direction. Also the strong winds, reported with colors shifting from blue to red in the wind rose, are all coming from the South-West quadrant. Telescope operations may be affected by windshake problems that causes the strucutre to vibrate and may affect normal and AO observations. By knowing that windshake effects are mostly pronounced when the wind speed flows frontally, with respect to the telescope mirrors, observations along lines of sight different from the South-West have a lower probability to incur on windshake events. The distribution of each year  is reported in Figures \ref{fig:wdall} (full wind rose), \ref{fig:wdsum} (summer wind rose) and \ref{fig:wdwin} (winter wind rose). In general we observe that summer has more variation on the wind direction, however this interests only the low wind speed (even the spike from North-West observable in 2015 and the one in 2018 to South-East corresponds mainly to wind speeds $<$ 5 m/s). Only the year 2020 saw relevant winds coming from East direction, with a non-negligible portion of winds in the range [10-20 m/s] which could be of relevance to telescope operations. Winter wind roses show a more consistent regime of winds coming almost exclusively from South-West.\\

\section{Conclusions}
\label{conclusions}
The correct characterization of the atmosphere and optical turbulence, i.e. climatology, is fundamental for all major ground-based telescopes assisted by complex adaptive optics systems. The LBT telescope is a unique telescope installation with cutting-edge adaptive optics and already assisted by atmospheric and OT forecasts from the ALTA project, which also makes use of a climatological description of the site for an accurate calibration of the Astro-Meso-NH code. In this contribution we presented preliminary results on the climatological analysis of LBT atmosphere coming from an action that is part of the ALTA Center project. The years analysed so far are from 2013 to 2021 (9 years). Products analysed are  measurements stored in the telemetry archive and produced by the monitors and instruments installed on the telescope dome. We observe that the site is stable with respect to OT, with a median seeing of 1.0 arcsec and a minimal seasonal variation. While interested by wet phenomena, the telescope site has also a 20\% of time characterized by a dry atmosphere, which makes it suitable to infrared observations. The strong winds ($\ge$ 15 m/s) are a relatively low occurrence ($\sim$ 10-15\% of the time), and mainly come from the South-West direction, thus observations planned on other directions should be relatively affected by the wind and the vibrations induced on the telescope structure.\\
In the future the telescope has already planned the installation of more instrumentation, such as a MASS, for measuring other relevant OT parameters such as $\tau_0$ and $\theta_0$, so a further more accurate characterization of its climatology will be available with positive impact on AO operations and ALTA forecasts.\\

\acknowledgments 
ALTA Center (ENV002) funded by LBTO, co-funded by the FCRF foundation action - N. 45103, Horizon 2020, Research and Innovation, and SOLARNET (N. 8241135).


\end{document}